\documentclass[12pt]{article}
\usepackage{a4wide,epsfig}
\catcode`@=11
\newcount\@tempcntc
\def\@citex[#1]#2{\if@filesw\immediate\write\@auxout{\string\citation{#2}}\fi
  \@tempcnta\z@\@tempcntb\m@ne\def\@citea{}\@cite{\@for\@citeb:=#2\do
    {\@ifundefined
       {b@\@citeb}{\@citeo\@tempcntb\m@ne\@citea\def\@citea{,}{\bf ?}\@warning
       {Citation `\@citeb' on page \thepage \space undefined}}%
    {\setbox\z@\hbox{\global\@tempcntc0\csname b@\@citeb\endcsname\relax}%
     \ifnum\@tempcntc=\z@ \@citeo\@tempcntb\m@ne
       \@citea\def\@citea{,}\hbox{\csname b@\@citeb\endcsname}%
     \else
      \advance\@tempcntb\@ne
      \ifnum\@tempcntb=\@tempcntc
      \else\advance\@tempcntb\m@ne\@citeo
      \@tempcnta\@tempcntc\@tempcntb\@tempcntc\fi\fi}}\@citeo}{#1}}
\def\@citeo{\ifnum\@tempcnta>\@tempcntb\else\@citea\def\@citea{,}%
  \ifnum\@tempcnta=\@tempcntb\the\@tempcnta\else
   {\advance\@tempcnta\@ne\ifnum\@tempcnta=\@tempcntb \else \def\@citea{--}\fi
    \advance\@tempcnta\m@ne\the\@tempcnta\@citea\the\@tempcntb}\fi\fi}
\catcode`@=12

\begin{document}
\newcommand{\be}{\begin{equation}}
\newcommand{\ee}{\end{equation}}
\newcommand{\bfm}[1]{\mbox{\boldmath$#1$}}
\newcommand{\bff}[1]{\mbox{\scriptsize\boldmath${#1}$}}

\newcommand{\al}{\alpha}
\newcommand{\bt}{\beta}
\newcommand{\lm}{\lambda}
\newcommand{\bea}{\begin{eqnarray}}
\newcommand{\eea}{\end{eqnarray}}
\newcommand{\gm}{\gamma}
\newcommand{\Gm}{\Gamma}
\newcommand{\dl}{\delta}
\newcommand{\Dl}{\Delta}
\newcommand{\ep}{\epsilon}
\newcommand{\vep}{\varepsilon}
\newcommand{\kp}{\kappa}
\newcommand{\Lm}{\Lambda}
\newcommand{\om}{\omega}
\newcommand{\pa}{\partial}
\newcommand{\nn}{\nonumber}
\newcommand{\dd}{\mbox{d}}
\newcommand{\grtsim}{\mbox{\raisebox{-3pt}{$\stackrel{>}{\sim}$}}}
\newcommand{\lessim}{\mbox{\raisebox{-3pt}{$\stackrel{<}{\sim}$}}}
\newcommand{\uk}{\underline{k}}
\newcommand{\gsim}{\;\rlap{\lower 3.5 pt \hbox{$\mathchar \sim$}} \raise 1pt \hbox {$>$}\;}
\newcommand{\lsim}{\;\rlap{\lower 3.5 pt \hbox{$\mathchar \sim$}} \raise 1pt \hbox {$<$}\;}

\title{
\begin{flushleft}
{\normalsize  TTP05-13\\[10mm]}
\end{flushleft}
Two-Loop Photonic Corrections to Massive Bhabha Scattering\\[2mm]}
\author{
  {\large A.A.~Penin} $^{a,b}$\\[2mm]
  $^a${\small {\it Institut f{\"u}r Theoretische Teilchenphysik,
  Universit{\"a}t Karlsruhe}}\\
  {\small {\it 76128 Karlsruhe, Germany}}\\[2mm]
  $^b${\small {\it Institute for Nuclear
  Research of Russian Academy   of Sciences,}}\\
{\small {\it  119899 Moscow, Russia}}
}
\date{}
\maketitle
\begin{abstract}
  We describe the details of the evaluation of the two-loop radiative
  photonic corrections to Bhabha scattering. The role of the corrections
  in the high-precision luminosity determination at present and future
  electron-positron colliders is discussed.
  \\[2mm]
\end{abstract}


\section{Introduction}
\label{int}
Electron-positron {\it Bhabha} scattering plays a special role in
particle phenomenology. It provides a very efficient tool for luminosity
determination at electron-positron colliders and thus it is crucial for
extracting physics from the experimental data.  Small angle Bhabha
scattering has been particularly effective as a luminosity monitor at
the energies of LEP and SLC because its cross section is large and QED
dominated \cite{Jad,MNP}.  At a future International Linear Collider
(ILC) the luminosity spectrum is not monochromatic due to beam-beam
effects.  Therefore measuring the cross section of the small angle
Bhabha scattering alone is not sufficient, and the acollinearity of the
large angle Bhabha scattering has been suggested for disentangling the
luminosity spectrum \cite{Too,Heu}.  Large angle Bhabha scattering is
important also at colliders operating at a center of mass energy
$\sqrt{s}$ of a few GeV, such as BABAR/PEP-II, BELLE/KEKB, BES/BEPC,
KLOE/DA$\Phi$NE, and VEPP-2M, where it is used to measure the luminosity
\cite{Car}. Since the accuracy of the theoretical evaluation of the
Bhabha cross section directly affects the luminosity determination,
remarkable efforts have been devoted to the study of the radiative
corrections to this process (see \cite{Jad} for an extensive list of
references).  Pure QED contributions are particularly important because
they dominate the radiative corrections to the large angle scattering at
intermediate energies 1-10~GeV and to the small angle scattering also at
higher energies.  The calculation of the QED radiative corrections to
the Bhabha cross section is among the classical problems of perturbative
quantum field theory with a long history.  The first order corrections
are well known (see \cite{Ber,CGR} and references therein).  To match
the impressive experimental accuracy the complete second order QED
effects have to be included on the theoretical side.  The evaluation of
the two-loop virtual corrections constitutes the main problem of the
second order analysis.  The complete two-loop virtual corrections to the
scattering amplitudes in the massless electron approximation have been
computed in Ref.~\cite{BDG}, where dimensional regularization has been
used for the infrared divergences.  However, this approximation is not
sufficient since one has to keep a nonvanishing electron mass to make
the result compatible with available Monte Carlo event generators
\cite{Jad,Car,Jad2,JPW,Arb4}.  Recently an important class of the second
order corrections, which include one closed fermion loop, has been
obtained for a finite electron mass \cite{BFMR} including the soft
photon bremsstrahlung \cite{Bon}.  A similar evaluation of the purely
photonic two-loop corrections is a challenging problem at the limit of
present computational techniques \cite{Smi,HeiSmi,Cza}. The most
complete result available so far can be found in Ref.~\cite{BonFer}
where the contribution of double box diagrams is still missing.  On the
other hand in the energy range under consideration only the leading
contribution in the small ratio $m_e^2/s$ is of phenomenological
relevance and should be retained in the theoretical evaluations.  For
arbitrary scattering angle even in this approximation only the two-loop
corrections enhanced by a power of the large logarithm $\ln(m_e^2/s)$
are known so far \cite{AKS,GTB}.  In the limit of the small scattering
angle, however, the structure of the corrections is much simpler
\cite{Fad2} that allowed for the evaluation of the corrections up to the
nonlogarithmic term \cite{Arb,Jad1}. The result for the nonlogarithmic
contribution for arbitrary scattering angle has been reported in a
letter \cite{Pen}.  It was obtained by employing the general theory of
infrared singularities in QED which allows to reduce the calculation in
the small electron mass approximation to the analysis of a strictly
massless scattering amplitude and the massive vector form factor. In the
present paper we describe the details of this calculation. In the next
section we outline the structure of the perturbative expansion for the
Bhabha cross section. In Sect.~\ref{sec3} we consider the structure of
the infrared logarithms and formulate the method of {\it infrared
  subtractions}. In Sect.~\ref{sec4} the explicit relation between the
amplitudes of the massive and massless Bhabha scattering is established
through the {\it infrared matching} procedure and the result for the
two-loop corrections to the massive Bhabha scattering is obtained.
Sect.~\ref{sec5} contains the numerical estimates and the summary.

\section{Perturbative expansion of the cross section}
\label{sec2}
We consider the phenomenologically interesting kinematical region
$s,~t,~u\gg m_e^2$, where all the terms suppressed by the electron mass
can be neglected.
The perturbative expansion for the Bhabha cross section in the
fine structure constant $\al$ is defined as follows
\be
\sigma=\sum_{n=0}^\infty\left({\al\over\pi}\right)^n\sigma^{(n)}\,.
\label{sigser}
\ee  
In the small electron mass approximation the leading
order differential cross section takes the form
\be
{{\rm d}\sigma^{(0)}\over{\rm d}\Omega}
={\al^2\over s}\left({1-x+x^2\over x}\right)^2+{\cal O}(m_e^2/s)\,,
\label{losig}
\ee 
where $x=(1-\cos\theta)/2$ and $\theta$ is the scattering angle.  

The virtual corrections taken separately suffer from the {\it soft}
divergences, which can be regulated {\it e.g.} by giving the photon a
small auxiliary mass $\lm$. These soft divergences are canceled in the
inclusive cross section when one adds the photonic bremsstrahlung
\cite{Kin,LeeNau}.  The standard approach to deal with the
bremsstrahlung is to split it into a soft part which accounts for the
emission of the photons with the energy below some cutoff $\vep_{cut}\ll
m_e$, and a hard part corresponding to the emission of the photons with
the energy above $\vep_{cut}$.  The infrared finite hard part is then
computed numerically using Monte-Carlo methods with physical cuts
dictated by the experimental setup.  At the same time the soft part is
computed analytically and combined with the virtual corrections ensuring
the cancellation of the singular dependence on $\lm$.  Note that in many
practical realizations of the Monte-Carlo event generators the
cancellation of the infrared singularities is build in and implemented
to high orders of perturbation theory for the amplitudes rather than for
the cross section (see {\it e.g.}  \cite{Jad2,JPW}). We will come back
to this issue in Sect.~\ref{sec4.2}.  Thus in the first order we
consider the sum of one-loop virtual correction and single soft photon
emission
\be
\delta^{(1)}=\delta^{(1)}_v+\delta^{(1)}_s\,,
\label{1ldec}
\ee
where 
\be
\delta^{(n)}\equiv{{\rm d}\sigma^{(n)}/{\rm d}\Omega
\over{\rm d}\sigma^{(0)}/{\rm d}\Omega}\,,
\label{decdef}
\ee
and the expressions for ${\delta^{(1)}_v}$ and ${\delta^{(1)}_s}$ are
given by Eqs.~(\ref{1lv},~\ref{1ls}) of the Appendix.  Eq.~(\ref{1ldec})
can be decomposed according to the asymptotic dependence on the electron
mass
\be
{{\rm d}\sigma^{(1)}\over{\rm d}\sigma^{(0)}}=
\delta^{(1)}_{1}\ln\left({s\over m_e^2}\right)
+\delta^{(1)}_{0}+{\cal O}(m_e^2/s)\,.
\label{1lexp}
\ee
For the  pure photonic correction the coefficients $\delta_i^{(1)}$ read
(see {\it e.g.} \cite{AKS,GTB})
\bea
\delta^{(1)}_{1}&=&4\ln\left({\vep_{cut}\over \vep}\right)+3\,,
\nn\\
&&
\nn\\
\delta^{(1)}_{0}&=&\left[-4+4\ln\left({x\over 1-x}\right)\right]
\ln\left({\vep_{cut}\over \vep}\right)-4-{2\over 3}\pi^2
-2{\rm Li}_2(x)+2{\rm Li}_2(1-x)+f(x)\,,
\label{1ldel}
\eea
where 
\bea
f(x)&=&(1-x+x^2)^{-2}\left\{\left({1\over 3}-{2\over 3}x
+{9\over 4}x^2-{13\over 6}x^3+{4\over 3} x^4\right)\pi^2+\left(3-4x
+{9\over 2}x^2-{3\over 2}x^3\right)\right.
\nn\\
&&\times\ln(x)+\left({3\over 4}x-{x^2\over 4}-{3\over 4}x^3+x^4\right)\ln^2(x)
+\left[-{1\over 2}x-{1\over 2}x^3+\left(2-4x+{7\over 2}x^2-x^3\right)\right.
\nn\\
&& 
\left.\times\ln(x)\bigg]\ln(1-x)+
\left(-1+{5\over 2}x-{7\over 2}x^2+{5\over 2}x^3-x^4\right)\ln^2(1-x)
\right\}\,,
\label{fx}
\eea
${\rm Li}_n(z)$ is the polylogarithm, $\vep=\sqrt{s}/2$, $\vep_{cut}$ is
the energy cut on the emitted soft photon.  The second order correction
can be represented as a sum of three terms
\be
\delta^{(2)}=\delta^{(2)}_{vv}
+\delta^{(2)}_{vs}+\delta^{(2)}_{ss}
\label{2ldec}
\ee
which correspond to the two-loop virtual correction including the
one-loop corrections to the amplitude square, one-loop
virtual correction to single soft photon emission, and the double
soft photon emission, respectively.  In the small electron mass limit 
it has the following decomposition
\be
\delta^{(2)}=
\delta^{(2)}_{2}\ln^2\left({s\over m_e^2}\right)
+\delta^{(2)}_{1}\ln\left({s\over m_e^2}\right)
+\delta^{(2)}_{0}+{\cal O}(m_e^2/s)\,.
\label{2lexp}
\ee 
The pure photonic, {\it i.e.} without closed fermion loops, 
logarithmically enhanced  contribution reads   
\cite{AKS}
\bea
\delta^{(2)}_{2}&=&8\ln^2\left({\vep_{cut}\over \vep}\right)+
12\ln\left({\vep_{cut}\over \vep}\right)+{9\over 2}\,,
\nn\\
\delta^{(2)}_{1}&=&\left[-16+16\ln\left({x\over 1-x}\right)\right]
\ln^2\left({\vep_{cut}\over \vep}\right)+\left[-28-{8\over 3}\pi^2
+12\ln\left({x\over 1-x}\right)-8{\rm Li}_2(x)\right.
\nn\\
&&
+8{\rm Li}_2(1-x)+4f(x)\bigg]\ln\left({\vep_{cut}\over\vep}\right)
-{93\over 8}-{5\over 2}\pi^2+6\zeta(3)-6{\rm Li}_2(x)
\nn\\
&&
+6{\rm Li}_2(1-x)+3f(x)\,,
\label{2ldel12}
\eea
where $\zeta(3)=1.202057\ldots$ is the value of the Riemann's
zeta-function. In the rest of the paper  we focus
on the  photonic contribution to  $\delta^{(2)}_{0}$.

\section{Structure of infrared logarithms}
\label{sec3}
The general problem of the calculation of the small electron mass
asymptotics of the corrections including the power-suppressed terms can
systematically be solved within the expansion by regions approach
\cite{BenSmi,Smi1}.  We, however, are interested only in the leading
order term.  The leading order contribution in Eq.~(\ref{2lexp})
contains the logarithmic terms, which become singular as $m_e$
approaches zero revealing the collinear divergences regulated by the
electron mass. In the massless limit both the collinear and the soft
divergences can be treated by dimensional regularization as well.  Here
we should note that the collinear divergences in the massless
approximation are also canceled in a cross section which is inclusive
with respect to real photons and electron-positron pairs collinear to
the initial or final state fermions \cite{SteWei}. This means that if an
angular cut on the collinear emission is sufficiently large,
$\theta_{cut}\gg \sqrt{m_e^2/s}$, the inclusive cross section is
insensitive to the electron mass and can in principle be computed with
$m_e=0$ by using dimensional regularization for the infrared divergences
for both virtual and real radiative corrections like it is done in the
theory of QCD jets.  However, as it has been mentioned above, all the
available Monte Carlo event generators for Bhabha scattering with
specific cuts on the photon bremsstrahlung dictated by the experimental
setup employ a nonzero electron mass as an infrared regulator, which
therefore has to be used also in the calculation of the virtual
corrections. As far as the leading term in the small electron mass
expansion is considered, the difference between the massive and the
dimensionally regularized massless Bhabha scattering can be viewed as a
difference between two regularization schemes for the infrared
divergences. With the known massless two-loop result at hand, the
calculation of the massive one is reduced to constructing the infrared
matching term which relates two above regularization schemes.  To
perform the matching we develop the method of infrared subtractions
which simplifies the calculation by fully exploiting the information on
the general structure of infrared singularities in QED.  The method was
originally applied in Ref.~\cite{FKPS} to the analysis of the two-loop
corrections to the vector form factor in an Abelian gauge model with
mass gap. Let ${\cal A}^{(2)}(m_e,\lm)$ be the two-loop contribution to
the massive electron-positron scattering amplitude with the photon mass
used to regulate the soft divergences.  The main idea of the method is
to construct an auxiliary amplitude $\bar{\cal A}^{(2)}(m_e,\lm)$, which
has the same structure of the infrared singularities but is sufficiently
simple to be evaluated at least in leading order in the small mass
expansion.  Then the difference ${\cal A}^{(2)}-\bar{\cal A}^{(2)}$ has
a finite limit $\delta{\cal A}^{(2)}$ as $m_e,~\lm$ tend to zero. This
quantity does not depend on the regularization scheme for ${\cal
  A}^{(2)}$ and $\bar{\cal A}^{(2)}$. It and can be evaluated by using
dimensional regularization for each term and then taking the limit of four
space-time dimensions.  The full amplitude is given by a sum
\be
{\cal  A}^{(2)}(m_e,\lm)= \bar{\cal A}^{(2)}(m_e,\lm)+
\delta{\cal  A}^{(2)}+{\cal O}(m_e,\lm)\,.
\label{split}
\ee 
Thus the infrared divergences, which induce the asymptotic dependence of
the virtual corrections on the electron and photon masses, are absorbed
into the auxiliary amplitude while the technically most nontrivial
calculation of the term $\delta{\cal A}^{(2)}$ is performed in the
massless approximation.  The matching of the massive and massless
results is necessary only for the singular auxiliary amplitude.  Note
that the method does not require a diagram-by-diagram subtraction of the
infrared divergences since only a general information on the infrared
structure of the total two-loop correction is necessary to construct
$\bar{\cal A}^{(2)}(m_e,\lm)$.  Our analysis is based on the following
infrared properties of the corrections to the scattering amplitudes:
\begin{itemize}
\item[(i)] exponentiation of the infrared logarithms 
           \cite{Sud,YFS,Jac,Mue,Col,Sen};
\item[(ii)] factorization of the collinear logarithms into external legs
           \cite{FreTay};
\item[(iii)] nonrenormalization of the infrared exponents \cite{YFS,Mue,Col}.
\end{itemize}
The first two properties are general and hold also for the closed
fermion loop corrections and non-Abelian gauge theories. The last
property is characteristic for the pure photonic corrections and plays a
crucial role in our analysis. In Sects.~\ref{sec3.1}-\ref{sec3.2} by
means of (i)-(iii) we show that $\bar{\cal A}^{(2)}(m_e,\lm)$ for the
photonic contributions can be constructed of the two-loop corrections to
the vector form factor and products of the one-loop corrections.

\subsection{Vector form factor}
\label{sec3.1}
The vector form factor ${\cal F}$ determines the electron scattering
amplitude in an external field. It plays a special role since it is the
simplest quantity which includes the complete information about the
collinear logarithms, which is directly applicable to a process with an
arbitrary number of electrons/positrons.  Let us consider three
different  {\it Sudakov} asymptotic regimes:
\begin{itemize}
\item[(a)] $m_e=\lm=0$; 
\item[(b)] $|Q|\gg m_e\gg\lm$;
\item[(c)] $|Q|\gg \lm \gg m_e$;
\end{itemize}
where $Q$ is the Euclidean momentum transfer.  In case (a) the soft
and collinear divergences are treated by dimensional regularization. In
case (b) the collinear and soft divergences are regularized by $m_e$
and $\lm$, respectively. In case (c) the photon mass regulates both
soft and collinear divergences.  Though (c) has no direct application to
QED, it is instructive to study yet another regularization scheme to get
deeper insight into the general structure of infrared logarithms. We
define the perturbative series for the form factor as follows: ${\cal
  F}=\sum_{n=0}^\infty\left({\al\over\pi}\right)^nf^{(n)}$.  The
one-loop coefficients read
\bea
f^{(1)}_a&=&\left[-{1\over 2\ep^2}-{3\over 4\ep}-2
+{\pi^2\over 24}+\left(-4+{\pi^2\over 16}+{7\over 6}\zeta(3)\right)\ep
+\left(-8+{\pi^2\over 6}+{7\over 4}\zeta(3)+{47 \over 2880}\pi^4\right)
\right.
\nn\\
&&\times\ep^2\bigg]\left({\mu^2\over Q^2}\right)^\ep \,,
\label{1lffa}\\
f^{(1)}_b&=&-{1\over 4}\ln^2\left({Q^2\over m_e^2}\right)+
\left[{1\over 2}\ln\left({\lm^2\over m_e^2}\right)+
{3\over 4}\right]\ln\left({Q^2\over m_e^2}\right)
-{1\over 2}\ln\left({\lm^2\over m_e^2}\right)-1+{\pi^2\over 12}
+{\cal O}(m^2_e,\lm^2)\,,
\nn\\
&&\label{1lffb}\\
f^{(1)}_c&=&-{1\over 4}\ln^2\left({Q^2\over \lm^2}\right)
+{3\over 4}\ln\left({Q^2\over \lm^2}\right)
-{7\over 8}-{\pi^2\over 6}+{\cal O}(\lm^2)\,.
\label{1lffc}
\eea
The asymptotic dependence of the form factor on $Q$ in the Sudakov limit
is governed by the evolution equation \cite{Mue,Col,Sen} which for the pure
photonic contribution takes the form
\be
{\partial\over\partial\ln\left(Q^2\right)}{\cal F}=
\left[ -{\al\over 2\pi}\ln\left(Q^2\right)
+\phi(m_e,\lm,\ep,\al)\right]{\cal F}\,,
\label{ffeq}
\ee
where the anomalous dimension $\phi(m_e,\lm,\ep,\al)$ is a series in
$\al$ with the coefficients depending on the infrared regulators.  We
can write down the solution of Eq.~(\ref{ffeq}) in the above three cases
\bea
{\cal F}_a&=&{\left(1+{\cal O}(\al)\right)}\exp \left\{-{\al\over 4\pi}
\left({2\over \ep^2}+{\left({3}+{\cal O}(\al)\right)}{1\over \ep}\right)
\left({\mu^2\over Q^2}\right)^\ep
\right\}\,,
\nn\\
{\cal F}_b&=&{\left(1+{\cal O}(\al)\right)}
\exp\left\{{\al\over 4\pi}\left[-\ln^2\left({Q^2\over m_e^2}\right)
+2\left[\ln\left({Q^2\over m_e^2}\right)-1\right]
\ln\left({\lm^2\over m_e^2}\right)+
\left(3+{\cal O}(\al)\right)\right.\right.
\nn\\
&&\left.\left.\times\ln\left({Q^2\over m_e^2}\right)\right]\right\}\,,
\nn\\
{\cal F}_c&=&{\left(1+{\cal O}(\al)\right)}
\exp \left\{{\al\over 4\pi}\left[-\ln^2\left({Q^2\over \lm^2}\right)
+{\left(3+{\cal O}(\al)\right)}
\ln\left({Q^2\over \lm^2}\right)\right]\right\}\,,
\label{ffsol}
\eea
where ${\cal O}(\al)$ indicates the presence of {\it all order}
corrections to the coefficients starting with ${\cal O}(\al)$ term.  In
derivation of Eq.~(\ref{ffeq}) we have taken into account that in the
case (b) also the logarithms of the photon mass exponentiate \cite{YFS}
\be
{\cal F}_b\propto\exp\left\{{\al\over 2\pi}
\left[\ln\left({Q^2\over m_e^2}\right)-1\right]
\ln\left({\lm^2\over m_e^2}\right)\right\}\,.
\label{lmexp}
\ee
The exponentiation of the ``Sudakov'' logarithms is a general
property valid also for the corrections due to the fermion loops and for
non-Abelian gauge theories. The exponent for the pure photonic
corrections, however, has two distinguished properties.  First, the
double logarithmic term in the exponent is protected against the
perturbative corrections. This fact is well known since the pioneering
works \cite{Sud,YFS,Mue,Col}.  A new observation is that beyond the
first order in $\al$ the coefficients of the series for the single
logarithmic term in the exponent are mass-independent and, therefore,
should be the same in all three cases under consideration.  The
derivation of this result to a large extent repeats the proof of the
nonrenormalization of the double logarithmic contribution.  We refrain
from giving the details of the derivation since it is not directly
related to the subject of the present paper.  We should emphasize that
the above properties are not valid for the nonphotonic corrections, {\it
  e.g.} the second order single logarithmic contribution of the closed
fermion loop to the exponent depends on the mass ratio $\lm/m_e$ and is
different for the cases (b) and (c) \cite{JKPS}.

Thus, as far as the photonic corrections are concerned, the perturbative
series for the logarithm of the form factor beyond  one loop includes
only the first power of the logarithm with the universal coefficients.
This means that the coefficients in the series for the form factor have
the following structure
\bea
f^{(2)}&=&{1\over 2}\left(f^{(1)}\right)^2+C^{(2)}\ln(Q^2)
+{\cal O}(\ln^0(Q^2))\,,
\label{2lff}\\
f^{(3)}&=&-{1\over 3}\left(f^{(1)}\right)^3
+f^{(2)}f^{(1)}+C^{(3)}\ln(Q^2)+{\cal O}(\ln^0(Q^2))\,,
\label{3lff}\\
\ldots\,,&&
\nn
\eea 
where each $C^{(n)}$, $n>1$ is equal for (a), (b), and (c).  This
prediction can be confronted with the explicit results for the two-loop
corrections which are available in all three cases and read
\bea
&&
\nn\\
f^{(2)}_a&=&{1\over 2}\left(f^{(1)}_a\right)^2-
\left({3\over 32}-{\pi^2\over 8}+{3\over 2}\zeta(3)\right){1\over 2\ep}
\left({\mu^2\over Q^2}\right)^{2\ep}
-{1\over 128}+{29\over 96}\pi^2-{15\over 8}\zeta(3)-{11\over 720}\pi^4\,,
\nn\\
&&\label{2lffa}\\
f^{(2)}_b&=&{1\over 2}\left(f^{(1)}_b\right)^2+\left({3\over 32}
-{\pi^2\over 8}+{3\over 2}\zeta(3)\right)
\ln\left({Q^2\over m_e^2}\right)+{11\over 8}+{17\over 32}\pi^2
-{9\over 4}\zeta(3)-{2\over 45}\pi^4
\nn\\
&&
-{\pi^2\ln(2)\over 2}\,,
\label{2lffb}\\
f^{(2)}_c&=&{1\over 2}\left(f^{(1)}_c\right)^2+
\left({3\over 32}-{\pi^2\over 8}+{3\over 2}\zeta(3)\right)
\ln\left({Q^2\over \lm^2}\right)+{51\over 128}+{15\over 16 }\pi^2
+5\zeta(3)-{83\over 360}\pi^4
\nn\\
&&-{2\over 3}\pi^2\ln^2(2)+{2\over 3}\ln^4(2)
+16\,{\rm Li}_4\left({1\over 2}\right)\,.
\label{2lffc}
\eea 
In the massless approximation the two-loop correction~(\ref{2lffa}) has
been known for a long time \cite{KraLam,MMN}.  The two-loop
correction~(\ref{2lffb}) was first obtained in Ref.\cite{Bur} by
integrating the dispersion relation with the spectral density computed
in Ref.~\cite{BMR}. The result has been checked in Ref.~\cite{Pen} and
can also be found in Ref.~\cite{MasRem} as a specific limit of the
result for an arbitrary momentum transfer.  The two-loop
correction~(\ref{2lffc}) has been obtained in Ref.~\cite{FKPS}.  As we
see the two-loop corrections indeed have the universal logarithmic term
corresponding to
\be
C^{(2)}={3\over 32}-{\pi^2\over 8}+{3\over 2}\zeta(3)\,.
\label{c2}
\ee 
By using the recent three-loop result \cite{MVV} for the massless case
we can completely predict the three-loop logarithmic corrections in
massive cases, which are given by Eq.~(\ref{3lff}) with
\be
C^{(3)}=-{29\over 2}-3\pi^2-68\zeta(3)+{16\over 3}\pi^2\zeta(3)+240\zeta(5)\,.
\label{c3}
\ee

\subsection{Scattering amplitude}
\label{sec3.2}
In the high energy limit the amplitude for the electron-positron
scattering has two components corresponding to the scattering of
particles of the same or opposite chirality.  We can write the
perturbative series for the amplitude as follows
\be
{\cal A}=\sum_{n=0}^\infty\left({\al\over\pi}\right)^n{\cal A}^{(n)}\,, 
\qquad {\cal A}^{(n)}=A^{(n)}{\cal A}^{(0)}\equiv 
\sum_{i=1}^2A^{(n)}_i{\cal A}^{(0)}_i\,,  
\qquad A^{(0)}_i=1\,,
\label{aser}
\ee
where ${\cal A}^{(0)}$ is a two component vector in the chiral basis
corresponding to the tree amplitude.  The collinear divergences
are completely determined  by external legs \cite{FreTay} and, therefore,
are the same for the scattering amplitude and the square of the form
factor.  It is convenient to introduce a reduced amplitude
\be
{\cal A}={\cal F}^2 \tilde{\cal A}\,,
\label{decom}
\ee 
which is free of collinear divergences. It satisfies a linear
differential equation \cite{Sen,Ste} which for the photonic contribution
takes the following form
\be
{\partial \over \partial \ln\left({Q^2}\right)} \tilde{\cal A}=
-{\al\over \pi} \ln\left({x\over 1-x}\right) \tilde{\cal A}\,,
\label{ameq}
\ee 
where the angular dependent anomalous dimension does not depend on
chirality. The solution of Eq.~(\ref{ameq}) reads
\bea
&&
\nn\\
\tilde{\cal A}|_{\lm=0}&=&
{\left({\cal A}^{(0)}+{\cal O}(\al)\right)}
\exp\left[{\al\over \pi} \ln\left({x\over 1-x}\right)
{1\over \ep}\left({\mu^2\over Q^2}\right)^\ep
\right]\,,
\nn \\
\tilde{\cal A}|_{\lm\ne0}&=&
{\left({\cal A}^{(0)}+{\cal O}(\al)\right)}
\exp\left[-{\al\over \pi} \ln\left({x\over 1-x}\right)
\ln\left({Q^2\over \lm^2}\right)\right]\,,
\label{amsol}
\eea 
where the corrections in the prefactor of the exponent are different for
different chiral components of the amplitude.  There are no photonic
corrections to the exponent in Eq.~(\ref{amsol}) \cite{YFS}.  Note that
in the case (b) all the singular dependence of the corrections to the
scattering amplitude on $m_e$ is absorbed into the form factor.
Eq.~(\ref{amsol}) implies that the logarithm of the reduced amplitude is
finite beyond one loop and the coefficients of the series for each
component of the chiral basis have the following structure
\bea
\tilde{A}^{(2)}_i&=&{1\over
  2}\left(\tilde{A}^{(1)}_i\right)^2+{\cal O}(\ln^0(Q^2))\,,
\label{2lam}\\
\tilde{A}^{(3)}_i&=&{1\over
  6}\left(\tilde{A}^{(1)}_i\right)^3+{\cal O}(\ln^0(Q^2))\,.
\label{3lam}\\
\ldots\,.&&
\nn
\eea
Now we can predict the singilar structure of the photonic corrections to
the full amplitude to all orders and to construct the auxilary amplitude
$\bar A^{(2)}$.  In the two-loop approximation by using
Eqs.~(\ref{2lff},~\ref{2lam}) we obtain
\be
\bar  A^{(2)}_i={1\over 2}\left(A^{(1)}_i\right)^2+
2\left[f^{(2)}-{1\over 2}\left({f^{(1)}}\right)^2\right]\,.
\label{auxam}
\ee
The expression~(\ref{auxam}) has all the necessary properties: it is
composed of the one-loop corrections to the chiral amplitudes and the
two-loop corrections to the form factor which are available in the
massive case and it has the same structure of infrared divergences as
the full amplitude.

\section{Two-loop corrections to the massive Bhabha cross section}
\label{sec4}
Once the result for the auxiliary amplitude in Eq.~(\ref{split}) is
known, the problem is to evaluate the difference
\be
\delta A^{(2)}_i= A^{(2)}_i-\left[{1\over 2}\left(A^{(1)}_i\right)^2+
2\left[f^{(2)}-{1\over 2}\left({f^{(1)}}\right)^2\right]\right]\,,
\label{match}
\ee 
which matches the auxiliary and the full amplitudes.  We should note
that though the different terms on the right hand side of
Eq.~(\ref{match}) taken separately are infrared divergent, their sum can
be transformed into convergent Feynman integrals.  In
similar way the pinch singularities disappear at the level of Feynman
integrals in the two-loop corrections to the static potential after the
proper infrared subtraction \cite{KPSSS}. Thus, in principle,
Eq.~(\ref{match}) does not need to be regularized. However, it is
simpler to take the available results for the different terms of
Eq.~(\ref{match}) in dimensional regularization and then to take the
limit $d\to 4$ in the sum.  The explicit result for the matrix elements
of the two-loop and the tree amplitudes in dimensional regularization
can be found in Ref.~\cite{BDG}.  To disentangle the infrared
divergences the authors of Ref.~\cite{BDG} used the formula suggested by
S. Catani in Ref.~\cite{Cat}. In the next section we describe how this
result can be matched to Eq.~(\ref{match}).

\subsection{Two-loop infrared matching}
\label{sec4.1}
In the case of pure photonic corrections the Catani formula for the
structure of the infrared divergences takes the following form
\bea
{\cal A}^{(1)}&=&{\bfm I}^{(1)}{\cal A}^{(0)}+{\cal A}^{(1)}_{\rm fin}\,,
\nn \\
{\cal A}^{(2)}&=&\left[-{1\over 2}\left({\bfm I}^{(1)}\right)^2
+{\bfm H}^{(2)}\right]{\cal A}^{(0)}+{\bfm I}^{(1)}{\cal A}^{(1)}
+{\cal A}^{(2)}_{\rm fin}\,,
\label{cateq}
\eea
where $A^{(n)}_{\rm fin}$ are finite in the limit $\ep\to 0$ and the
infrared divergences are described by the operators
\bea
{\bfm I}^{(1)}&=&{e^{-\ep\gamma_E}\over \Gamma(1-\ep)}\left({1\over
    \ep^2}+{3\over 2\ep}\right)\left[-\left({\mu^2\over -s}\right)^\ep-
\left({\mu^2\over -t}\right)^\ep+\left({\mu^2\over -u}\right)^\ep\right]\,,
\nn \\
{\bfm H}^{(2)}&=&{e^{-\ep\gamma_E}\over \Gamma(1-\ep)}{1\over \ep}\left(
{3\over 32}-{\pi^2\over 8}+{3\over 2}\zeta(3) \right)
\left[-\left({\mu^2\over -s}\right)^\ep-
\left({\mu^2\over -t}\right)^\ep+\left({\mu^2\over -u}\right)^\ep\right]\,.
\label{catop}
\eea 
which are diagonal in the chiral basis.  The form of the nonsingular
terms in Eq.~(\ref{catop}) is a matter of convention. We use the one of
Refs.~\cite{BDG,Cat}\footnote{Our normalization of the operators differs
  from \cite{BDG,Cat} by the overall factor 2 per loop.}. It is easy to
check that the above expression is in full agreement with
Eq.~(\ref{auxam}).  Indeed, Eq.~(\ref{cateq}) is invariant under a
redefinition
\bea
{\bfm I}'^{(1)}&=&{\bfm I}^{(1)}+G\,, \qquad
{A'}^{(1)}_{\rm fin}={A}^{(1)}_{\rm fin}-G \,, 
\nn\\
{\bfm H}'^{(2)}&=&{\bfm H}^{(2)}+F\,,\qquad\hspace{-6pt}
{A'}^{(2)}_{\rm fin}=A^{(2)}_{\rm  fin}-\left({1\over 2}G^2+F\right)\,,
\label{chop}
\eea
where $G$ and $F$ are the two-component functions of $x$ and $\ep$ which
are regular at $\ep=0$. By choosing
\be
G=A^{(1)}_{\rm fin}\,, \qquad F= 2f^{(2)}-(f^{(1)})^2-{\bfm H}^{(2)}
\label{gf}
\ee 
we reproduce the structure of Eqs.~(\ref{auxam},~\ref{match}) with 
\be
A^{(1)}={\bfm I}'^{(1)}\,, \qquad  2f^{(2)}-(f^{(1)})^2={\bfm H}'^{(2)}\,
\qquad   \delta A^{(2)}={A'}^{(2)}_{\rm fin}\,.
\label{newop}
\ee 
Thus Eq.~(\ref{cateq}) can be considered as a direct consequence of the
evolution equations~(\ref{ffeq},~\ref{ameq}).  Moreover, by analysing
the evolution equations we can predict the form of the operator ${\bfm
  H}^{(2)}$ for the scattering amplitude which was not determined in
Ref.~\cite{Cat} but instead has been found by explicit calculation
\cite{BDG}.  In fact the same is true for the four-quark amplitude where
the Catani formula is a direct concequence of the non-Abelian evolution
equations \cite{Sen}.

By using Eq.~(\ref{gf}) one can transform the result of Ref.~\cite{BDG}
for the matrix element of ${\cal A}^{(2)}_{\rm fin}$ into the one of
$\delta {\cal A}^{(2)}$.  However, we prefer to use directly the result
of \cite{BDG} for the finite part of the two-loop corrections and find
the expressions for the operators ${\bfm I}^{(1)}$ and ${\bfm H}^{(2)}$
corresponding to the mass regularization of the infrared divergences.
To perform this infrared matching let us first note that the primed
operators defined through Eqs.~(\ref{chop},~\ref{gf}) are given by the
sum of Feynman integrals corresponding to the one-loop correction to the
amplitude and two-loop correction to the logarithm of the form factor,
respectively.  Therefore, in contrast to the original
definition~(\ref{catop}), it is straightforward to obtain the variation
of the primed operators with the change of the infrared regularization.
For ${\bfm I}'^{(1)}$ it consists in replacing the one-loop massless
result by the massive one. For ${\bfm H}'^{(2)}$ the matching term which
relates the dimensionally regularized and the massive result is given by
twice the difference of the nonlogarithmic terms of Eq.~(\ref{2lffb})
and Eq.~(\ref{2lffa}), where the $(f^{(1)})^2/2$ contribution is
subtracted.  As we have already pointed out, the finite part
${A'}^{(2)}_{\rm fin}= \delta A^{(2)}$ does not depend on the
regularization. Now we can perform the inverse transformation to the
operators ${\bfm I}^{(1)}$ and ${\bfm H}^{(2)}$.  Note that we are
interested in the limit $d=4$ and only need the value of the functions
$F$ and $G$ at $\ep=0$. The finite part of the two-loop correction to
the amplitude ${A}^{(2)}_{\rm fin}$ does not change after these
transformations while ${\bfm I}^{(1)}$ and ${\bfm H}^{(2)}$ become
logarithmic functions of electron and photon masses
\bea 
{\bfm I}^{(1)}&=&-{1\over 2}\ln^2\left({s\over m_e^2}\right)+ 
\left[\ln\left({\lm^2\over m_e^2}\right)+ 
{3\over 2}-\ln\left({x\over 1-x}\right)+i\pi\right]
\ln\left({s\over m_e^2}\right) +\bigg[-1
\nn \\
&& 
\left.+\ln\left({x\over 1-x}\right)-i\pi\right] 
\ln\left({\lm^2\over m_e^2}\right)+2 -{2\over 3}\pi^2
+{3\over 2}\ln\left({x\over 1-x}\right) -{1\over 2}\ln^2(x)
\nn \\
&& 
+{1\over 2}\ln^2(1-x)-{3\over 2}i\pi\,,
\nn \\
{\bfm H}^{(2)}&=&\left( {3\over 16}-{\pi^2\over 4}+3\zeta(3)
\right)\left[\ln\left({s\over m_e^2}\right)+\ln\left({x\over 1-x}\right)
-i\pi\right]+{177\over 64}+{11\over 24}\pi^2
\nn \\
&&
-{3\over 4}\zeta(3)-{7\over 120}\pi^4 -\pi^2\ln(2)\,.
\label{myop}
\eea 
By plugging Eq.~(\ref{myop}) into Eq.~(\ref{cateq}) we reproduce the
known result for the logarithmic corrections to the amplitudes.  Beside
the logarithmic correction the operators~(\ref{myop}) produce a
nonlogarithmic contribution in Eq.~(\ref{cateq}) which can be considered
as the matching term between the dimensionally regularized and the
massive result for the amplitudes.  Note that in Ref.~\cite{BDG} the
explicit result is given only for the matrix elements of the two-loop
and the tree amplitudes rather than the expressions for the chiral
amplitudes.  This, however, is sufficient for the matching because the
operators Eq.~(\ref{catop}) are diagonal in the chiral basis.

\subsection{The result}
\label{sec4.2}
Now we are in a position to derive the result for the second order
correction to the cross section of the massive Bhabha scattering. It can
be split into three parts:
\begin{itemize}
\item[(i)] the corrections involving the soft real emission;
\item[(ii)] the interference of the one-loop corrections to the
  amplitudes;
\item[(iii)] the interference of the two-loop corrections and the tree
  amplitudes.
\end{itemize}
The soft photon emission is known to factorize and the corresponding
second order corrections to the cross section introduced in
Eq.~(\ref{2ldec}) are of the following form
\be
\delta^{(2)}_{vs}=\delta^{(1)}_v\delta^{(1)}_s\,, 
\qquad
\delta^{(2)}_{ss}={1\over 2}{\delta^{(1)}_s}^2\,.
\label{2lsvss}
\ee 
With the known one-loop corrections to the chiral amplitudes at hand
(see {\it e.g.} Ref.~\cite{Boh}) it is straightforward to obtain the
corresponding interference term in the cross section, Eq.~(\ref{inter})
of the Appendix. The derivation of the contribution (iii) has been
described in the previous section.  Collecting all the contributions we
obtain the result for the nonlogarithmic photonic correction which is
given by Eq.~(\ref{result}) of the Appendix.

In the limit of small scattering angles the virtual corrections to the
cross section are completely determined by the corrections to the
electron and positron form factors in the $t$-channel amplitude
\cite{Fad2}.  We check that our result for the the virtual
corrections in the the limit $x\to 0$ reduces to
\be
\left.\delta^{(2)}_{vv}\right|_{x\to 0}=6\left(f^{(1)}_b\right)^2
+4f^{(2)}_b+{\cal O}(x)\,,
\label{smallxf}
\ee
where $f^{(n)}_b$ are given by Eqs.~(\ref{1lffb},~\ref{2lffb}) with
$Q^2=xs$. This agrees with the asymptotic small angle expression given
in \cite{Arb} which is quite a nontrivial check of our result. 
The two-loop nonlogarithmic corrections to the cross section
in the small angle limit becomes
\bea
\left.\delta^{(2)}_{0}\right|_{x\to 0}&=&
\left[8\ln^2\left({\vep_{cut}\over \vep}\right)
+12\ln\left({\vep_{cut}\over \vep}\right)+{9\over 2}\right]\ln^2(x)
+\Bigg[-16\ln^2\left({\vep_{cut}\over \vep}\right)
-28\ln\left({\vep_{cut}\over \vep}\right)
\nn\\
&&\left.-{93\over 8}-{\pi^2\over 2}+6\zeta(3)\right]\ln(x)
+8\ln^2\left({\vep_{cut}\over \vep}\right)
+16\ln\left({\vep_{cut}\over \vep}\right)+C\,,
\label{smallxd}
\eea
where
\be
C={27\over 2}+{17\over 8}\pi^2-9\zeta(3)-{8\over 45}\pi^4
-2\pi^2\ln(2)=-0.744199\ldots\,.
\label{c}
\ee 
Note that our result is not valid for very small scattering angles
corresponding to $x\lessim m_e^2/s$ and for almost backward scattering
corresponding to $1-x\lessim m_e^2/s$, where the power-suppressed terms
of the form $m_e^2/t$ and $m_e^2/u$ become important.

Our result should be combined with the Monte-Carlo evaluation of the
hard brems\-strahlung.  As it has already been mentioned in many practical
realizations of the Monte-Carlo event generators the cancellation of
infrared divergences between virtual and soft real corrections is
implemented to high orders in perturbation theory by using the
exponentiation property discussed above. In this case the Monte-Carlo
result for the cross section already includes a part of the second order
virtual and soft real corrections.  This part depends on the specific
realization of the Monte-Carlo algorithm and should be subtracted from
the result of the paper.

\section{Numerical estimates and summary}
\label{sec5}
 
\begin{figure}
\begin{tabular}{cc}
\hspace*{-25pt}\epsfig{figure=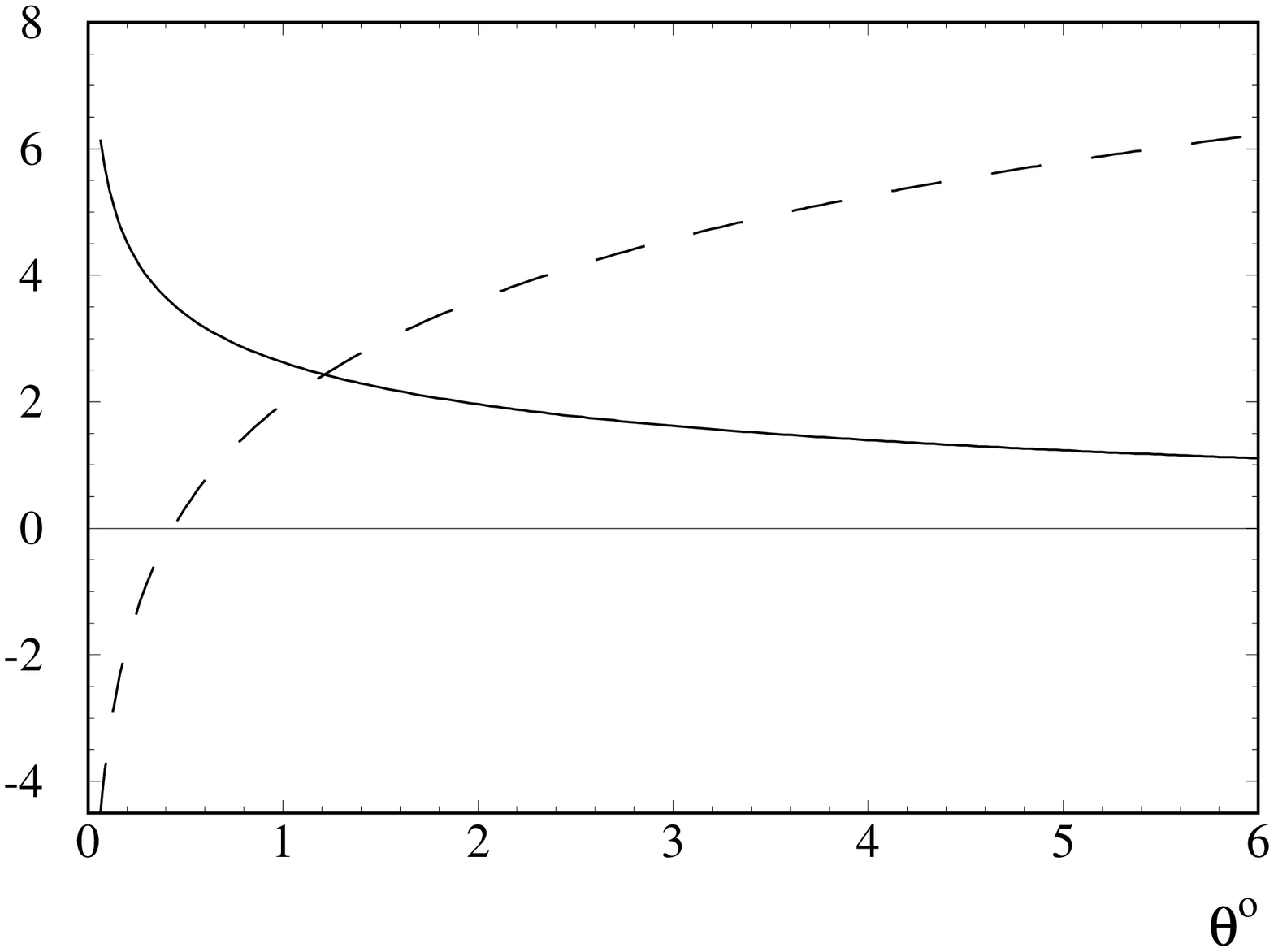,height=6cm}&
\hspace*{-25pt}\epsfig{figure=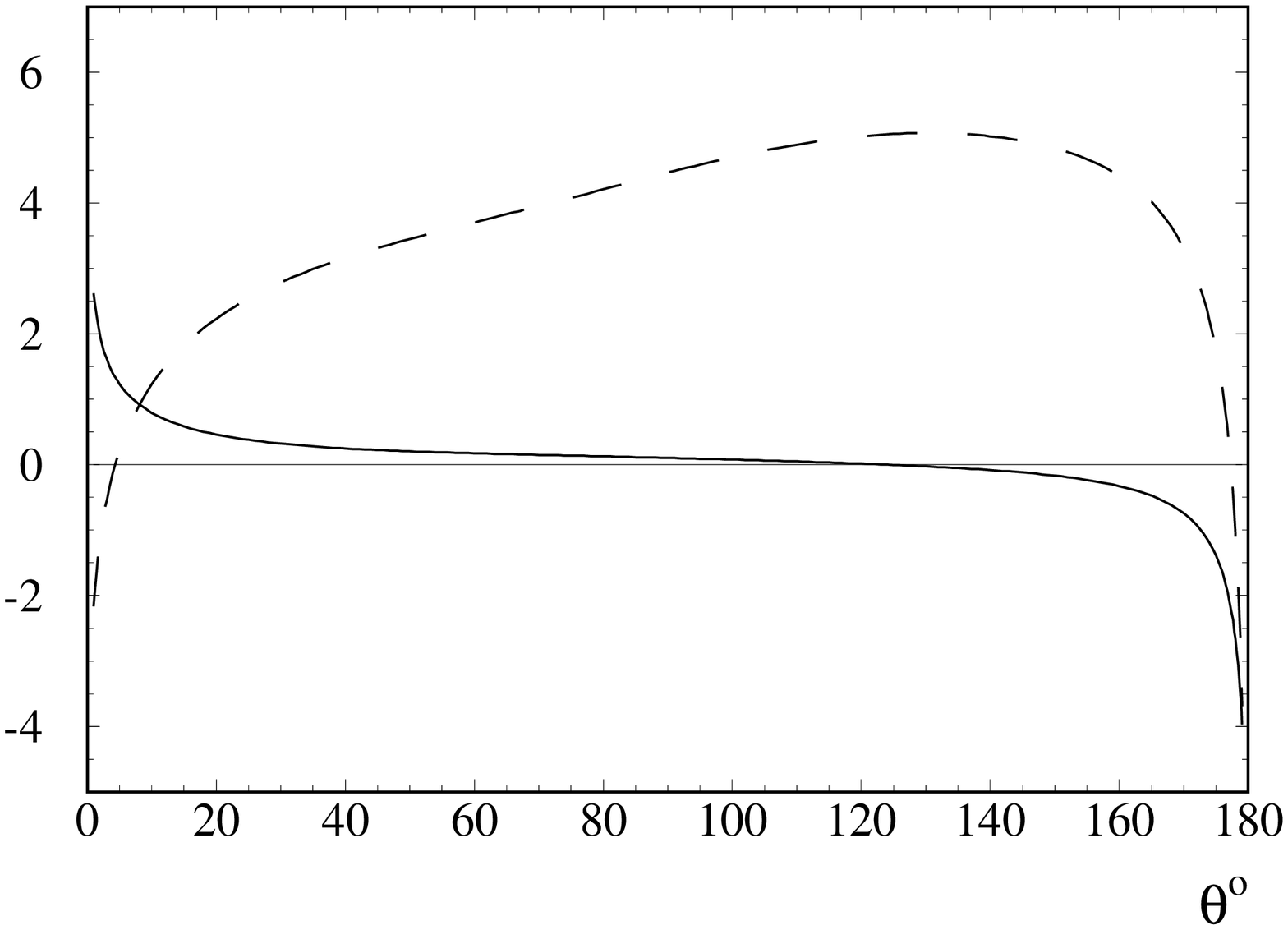,height=6cm}\\
\hspace*{-25pt} $(a)$ & 
\hspace*{-25pt} $(b)$
\end{tabular}
\caption{\label{fig1} \small
  $(a)$ Logarithmically enhanced (dashed line) and nonlogarithmic (solid
  line) second order corrections to the differential cross section of
  the small angle Bhabha scattering as functions of the scattering angle
  for $\sqrt{s}=100$~GeV and $\ln(\vep_{cut}/\vep)=0$, in permill.
  $(b)$ The same as (a) but for the large angle Bhabha scattering and
  $\sqrt{s}=1$~GeV.}
\end{figure}

Let us now discuss the phenomenological relevance of our result.  The
determination of the luminosity for the GigaZ option of ILC is most
demanding with respect to the theoretical predictions for the small
angle Bhabha scattering. It requires the accuracy at the level of 0.1
permill \cite{Heu}.  At the same time the low-energy experiments aimed
at the determination of the hadronic vacuum polarization contribution
through the measurement of $\sigma(e^+e^-\to~{\rm hadrons})$ require
about one permill accuracy of the large angle Bhabha cross section.
Such a high accuracy is necessary to reduce theoretical uncertainty due
to the hadronic vacuum polarization contribution to the muon anomalous
magnetic moment and to the value of the QED coupling constant at $Z$
peak (see {\it e.g.} Ref.~\cite{Jeg}). Neither of the existing
Monte-Carlo event generators for small \cite{Jad,MNP,Jad1} and large angle
\cite{Car,JPW,Arb4} Bhabha scattering, which as yet do not incorporate
the complete second order QED corrections, can guarantee the required
precision.

The second order photonic corrections to the differential cross section
$(\al/\pi)^2{\rm d}\sigma^{(2)}/{\rm d}\sigma^{(0)}$ are plotted as
functions of the scattering angle for the small angle Bhabha scattering
at $\sqrt{s}=100$~GeV on Fig.~(\ref{fig1}a) and for the large angle
Bhabha scattering at $\sqrt{s}=1$~GeV on Fig.~(\ref{fig1}b).  We
separate the logarithmically enhanced corrections given by the first two
terms of Eq.~(\ref{2lexp}) and the nonlogarithmic contribution given by
the last term of this equation.  All the terms involving a power of the
logarithm $\ln(\vep_{cut}/\vep)$ are excluded from the numerical
estimates because the corresponding contribution critically depends on
the event selection algorithm and cannot be unambiguously estimated
without imposing specific cuts on the photon bremsstrahlung.  The actual
impact of the two-loop virtual corrections on the theoretical
predictions can be determined only after the result of the paper is
consistently implemented into the Monte-Carlo event generators.
Nevertheless, the above na\"ive procedure can be used to get a rough
estimate of the magnitude and the structure of the corrections.  We
observe that for scattering angles $\theta \lessim 18^o$ and $\theta
\grtsim 166^o$ the nonlogarithmic contribution exceeds a benchmark of
$0.5$ permill which makes it relevant for the luminosity determination
at the low-energy electron-positron colliders.  For the small scattering
angles the second order correction reaches a few permill in magnitude.
Here we should note that BHLUMI event generator \cite{Jad1} used for
luminosity determination at LEP includes the total leading logarithmic
second order contribution enhanced by the factor $\ln^2(t/m^2_e)$ as
well as the bulk of the subleading contribution.  In fact the remaining
part of the subleading photonic corrections to the cross section has
been computed in the small-angle approximation \cite{Jad2} but has not
been included in the code.  According to Ref.~\cite{Jad2} this missing
correction amounts for approximately $0.14$ permill for the energy and
scattering angles characteristic to LEP, which is relevant for the GigaZ
accuracy.

\begin{figure}
\begin{center}
\begin{tabular}{cc}
\hspace*{-25pt}\epsfig{figure=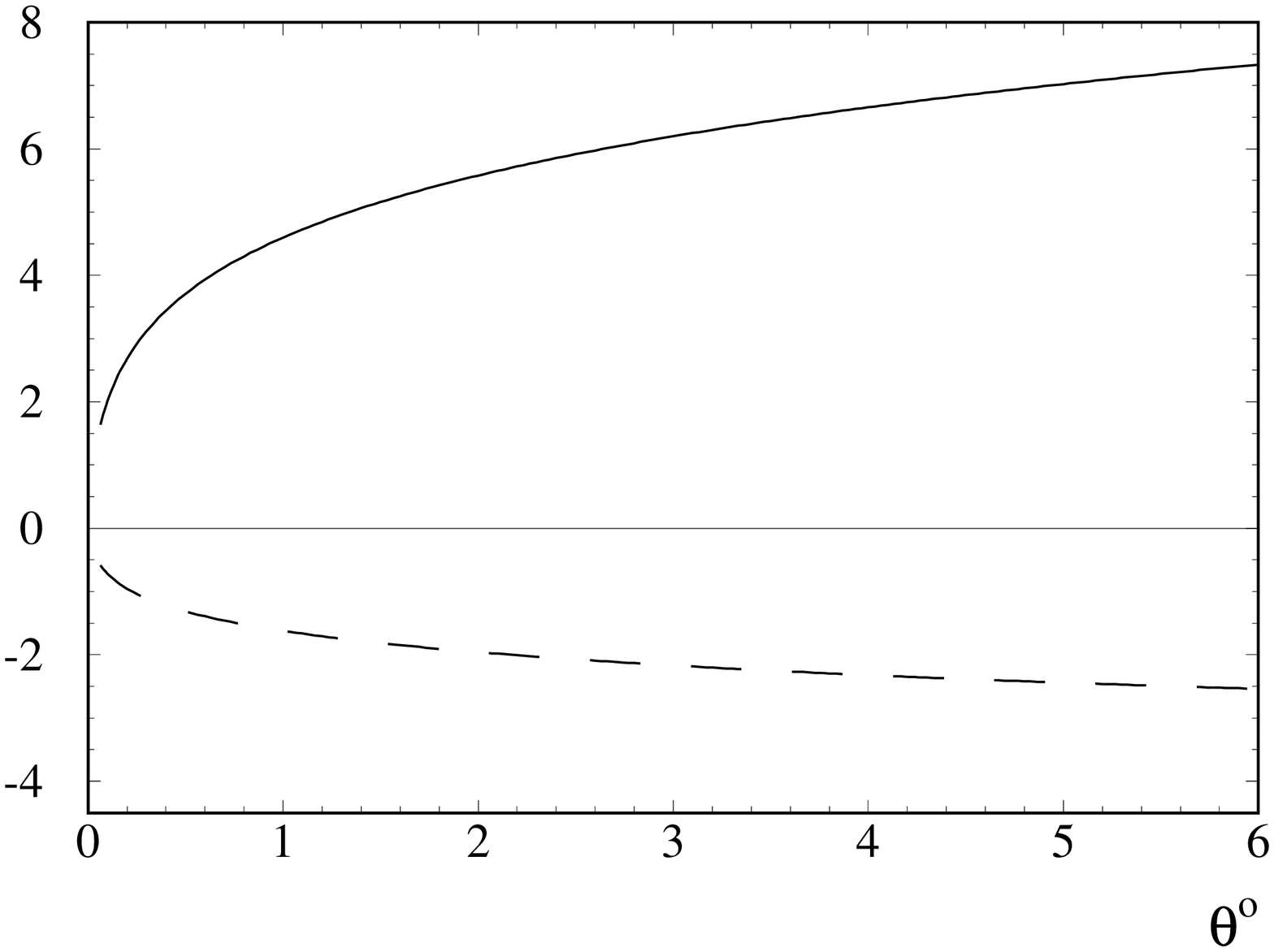,height=6.0cm}&
\hspace*{-25pt}\epsfig{figure=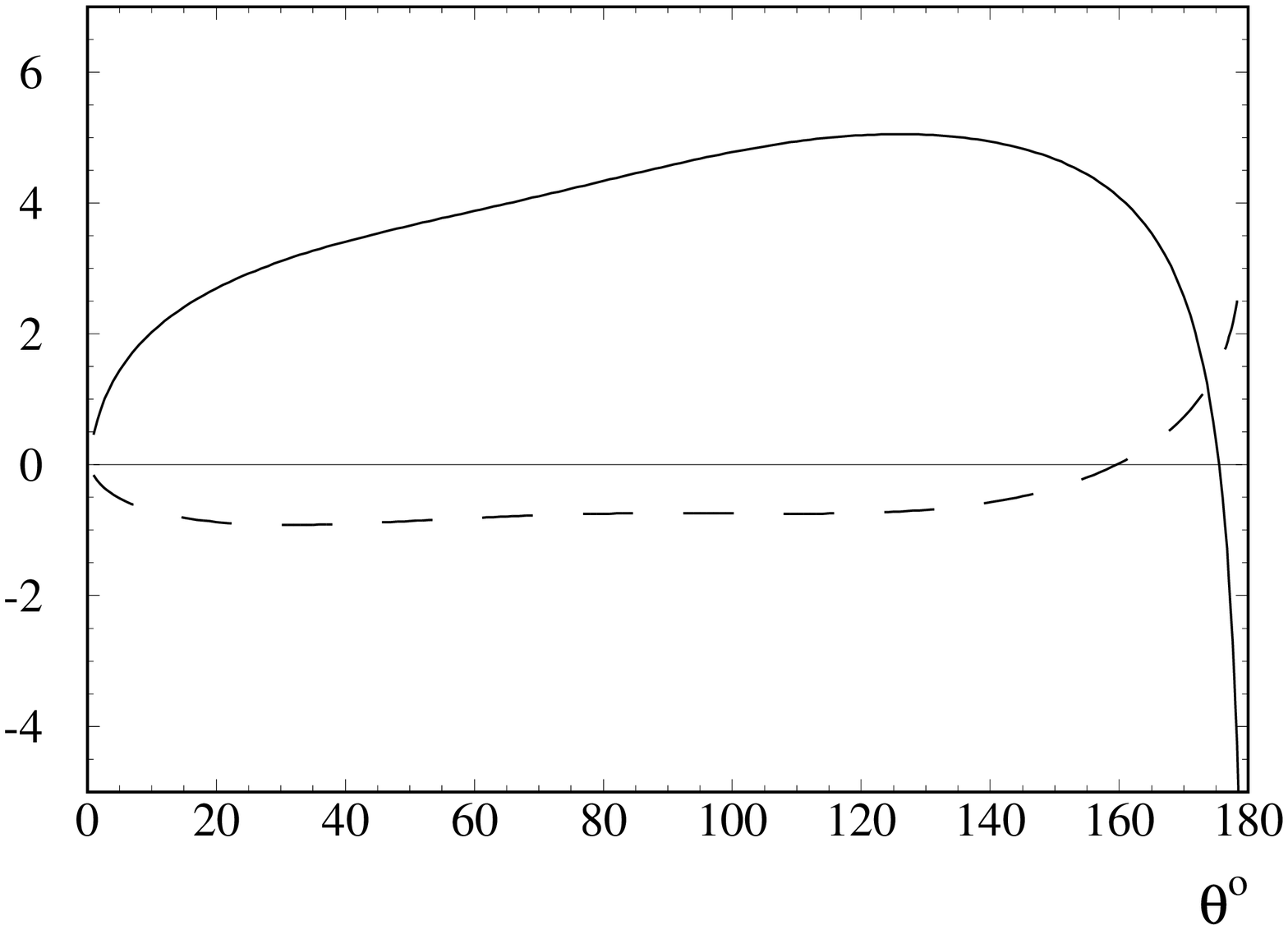,height=6.0cm}\\
\hspace*{-25pt} $(a)$ & 
\hspace*{-25pt} $(b)$
\end{tabular}
\caption{\label{fig2} \small
  $(a)$ Photonic (solid line) and  fermionic (dashed line) 
  second order corrections to the differential cross
  section of the small angle Bhabha scattering as functions of the
  scattering angle for $\sqrt{s}=100$~GeV and
  $\ln(\vep_{cut}/\vep)=\ln(\vep^{e^+e^-}_{cut}/\vep)=0$, in permill.
  $(b)$ The same as $(a)$ but for the large angle Bhabha scattering and
  $\sqrt{s}=1$~GeV. }
\end{center}
\end{figure}

To get the total second order correction (without the hard
bremsstrahlung) our result should be combined with the fermionic
contribution. In Fig.~(\ref{fig2}) we plot the second order photonic
contribution against the fermionic one in the case of one light flavor.
The fermionic contribution incorporates the second order corrections
with one closed fermion loop including the single soft photon emission
\cite{Bon} and the contribution due to the emission of the soft real
electron-positron pair of the energy below a cutoff
$\vep^{e^+e^-}_{cut}\ll s$.  The latter has been computed in
Ref.~\cite{Arb1} in the logarithmic approximation and cancels the
artificial $\ln^3(m_e^2/s)$ term of the closed fermion loop
contribution. Note that we do not include a trivial contribution with
two closed fermion loops which can be taken into account through the
one-loop renormalization group running of $\al$ in the tree amplitudes.

Finally, we would like to mention the electroweak corrections to Bhabha
scattering which can be important at the considered level of accuracy.
The one-loop correction is well known \cite{Boh}.  For the large angle
scattering above the electroweak scale, however, the two-loop
electroweak corrections could be important.  In the case of
$e^+e^-\to\mu^+\mu^-$ annihilation the corrections enhanced in the high
energy limit by a power of the large logarithm $\ln(M^2/s)$, where $M$
stands for $W$ or $Z$ boson mass, have been computed in
\cite{JKPS,Fad,KPS,KMPS}.  They dominate the electroweak corrections for
the energies $\sqrt{s}\grtsim 500$~GeV characteristic to ILC.  Due to
the strong numerical cancellations between the terms with different
powers of the large logarithm the total electroweak logarithmic two-loop
contribution does not exceed a few permill in this energy region.  This
analysis can be generalized to the large angle Bhabha scattering by
adding the $t$-channel contribution.

To conclude, we have derived the two-loop radiative photonic corrections
to Bhabha scattering in the leading order of the small electron mass
expansion up to nonlogarithmic term.  Together with the result of
Ref.~\cite{BFMR,Bon} for the fermion loop corrections our result gives a
complete expression for the two-loop virtual corrections. It should be
incorporated into the Monte Carlo event generators to match the demands
of the present and future electron-positron colliders for the accuracy
of the luminosity determination.

\vspace{5mm}
\noindent
{\bf Acknowledgments}\\[2mm]
I would like to thank V.A. Smirnov for his help in the evaluation of the
two-loop correction to the form factor and O.L. Veretin for pointing out
that the method of Ref.~\cite{FKPS} can be applied to the analysis of
Bhabha scattering.  I am grateful to R. Bonciani and A. Ferroglia
\cite{BonFer} for cross-checking a large part of the result and for
sharing the code for the fermion loop corrections.  I thank J.H. K\"uhn
and M.  Steinhauser for carefully reading the manuscript and useful
comments.  The work was supported in part by BMBF Grant No.\ 05HT4VKA/3
and Sonderforschungsbereich Transregio 9.

\section*{Appendix}
The one-loop virtual photonic correction to the normalized cross section
reads
\bea
\delta^{(1)}_v&=&
-\ln^2\left({s\over m_e^2}\right)+\left[2\ln\left({\lm^2\over m_e^2}\right)
+3-2\ln\left({x\over 1-x}\right)\right]
\ln\left({s\over m_e^2}\right)+
\left[-2+2\ln\left({x\over 1-x}\right)\right]
\nn \\
&&
\times\ln\left({\lm^2\over m_e^2}\right)
-4 -\ln^2(x)+\ln^2(1-x)+f(x)\,,
\label{1lv}
\eea 
where $f(x)$ is given by Eq.~(\ref{fx}).
The correction due to the single soft photon emission reads
\bea
\delta^{(1)}_s&=&
\ln^2\left({s\over m_e^2}\right)+\left[
4\ln\left({\vep_{cut}\over\vep}\right)-2\ln\left({\lm^2\over m_e^2}\right)+2\ln\left({x\over 1-x}\right)\right]
\ln\left({s\over m_e^2}\right)+
\bigg[-2
\nn \\
&&
\left.+2\ln\left({x\over 1-x}\right)\right]\left[
2\ln\left({\vep_{cut}\over\vep}\right)-\ln\left({\lm^2\over m_e^2}\right)\right]
-{2\over 3}\pi^2 +\ln^2(x)-\ln^2(1-x)
\nn \\
&&-2{\rm Li_2}(x)+2{\rm Li_2}(1-x)\,.
\label{1ls}
\eea
The second order contribution to the cross section due to the
interference of the one-loop virtual corrections to the amplitude reads
{\small
\bea
\lefteqn{\delta^{(1\times 1)}_{vv}=
4+\left({1-x+x^2}\right)^{-2}\Bigg(
\left(-{2\over 3}+{4\over 3}x-{13\over 4}x^2+{17\over 6}x^3
-{5\over 12}x^4\right)\pi^2+\left({1\over 36}- {x\over 18} + 
{13\over 24}x^2-{49\over 72}x^3\right.
\nn} \\
&&
\left. +
{137\over 288}x^4\right)\pi^4
+\left[-6+8x-9x^2+3x^3
+\left({1\over 2}+{5\over 6}x-{x^2\over 8}
+{5\over 8}x^3-3x^4\right)\pi^2\right]\ln(x)
+\Bigg[{17\over 4}-7x
\nn \\
&&
\left.+{31\over 4}x^2
-{5\over 2}x^3+\left({5\over 6}-{x\over 24}
+{x^2\over 12}+{x^3\over 12}+{17\over 16}x^4
\right)\pi^2\right]\ln^2(x)
+\left(-{3\over 2}+{25\over 8}x-2x^2-{x^3\over 8}\right)\ln^3(x)
\nn \\
&&
+\left({1\over 4}-{7\over 8}x+{33\over 32}x^2
-{x^3\over 8}
+{x^4\over 32}\right)\ln^4(x)
+\Bigg\{x+x^3+\left({x\over 6}+{3\over 2}x^2
-{101\over 24}x^3+3x^4\right)\pi^2
+\bigg[-4
\nn \\
&&
\left.+{29\over 4}x
-{29\over 4}x^2+2x^3
+\left({1\over 3}+{x\over 3}-{7\over 6}x^2
+{7\over 3}x^3-{17\over 8}x^4\right)\pi^2\right]\ln(x)
+\left(3-{15\over 4}x+{3\over 2}x^2
+{3\over 8}x^3\right)
\nn \\
&&
\times\ln^2(x)+\left(-1+{11\over 4}x-{9\over 4}x^2+{x^3\over 2}
-{x^4\over 8}\right)\ln^3(x)\Bigg\}\ln(1-x)
+\bigg[-x+{5\over 4}x^2-x^3
+\left({1\over 8}-{5\over 12}x
\right.
\nn \\
&&
\left.+{37\over 24}x^2
-{23\over 12}x^3+{17\over 16}x^4\right)\pi^2
+\left({x\over 4}+{3\over 8}x^2
-{3\over 8}x^3\right)\ln(x)
+\left(1-{9\over 4}x+{15\over 8}x^2
-{3\over 4}x^3
+{3\over 16}x^4\right)
\nn \\
&&
\times\ln^2(x)\bigg]\ln^2(1-x)
+\left[{x\over 8}-{x^2\over 2}+{x^3\over 8}
+\left({x\over 2}-{3\over 4}x^2+{x^3\over 2}
-{x^4\over 8}\right)\ln(x)\right]\ln^3(1-x)
+\left({1\over 32}\right.
\nn \\
&&
\left.-{x\over 8}+{x^2\over 4}
-{x^3\over 8}+{x^4\over 32}\right)
\ln^4\left(1-x\right)\,,
\label{inter}
\eea
}
where the trivial terms proportional to $\ln(\lm^2/m_e^2)$ and 
$\ln(m_e^2/s)$ are omitted. 

The total second order nonlogarithmic photonic contribution 
to the normalized cross section reads
{\small
\bea
\lefteqn{\delta^{(2)}_{0}=
8{\cal L}_{\vep}^2
+\left({1-x+x^2}\right)^{-2}\bigg[\left({4\over 3}
-{8\over 3}x-x^2+{10\over 3}x^3-{8\over 3}x^4\right)\pi^2
+\left(-12+16x-18x^2+6x^3\right)\ln(x)
\nn }\\
&&
+\left(2x+2x^3\right)\ln(1-x)+\left(-3x+x^2+3x^3-4x^4\right)
\ln^2(x)+\left(-8+16x-14x^2+4x^3\right)\ln(x)
\nn \\
&&
\times\ln(1-x)+\left(4-10x+14x^2-10x^3+4x^4\right)\ln^2(1-x)
+\left({1-x+x^2}\right)^2(16+8{\rm Li_2}(x)
\nn \\
&&
-8{\rm Li_2}(1-x))
\bigg]{\cal L}_{\vep}
+{27\over 2}-2\pi^2\ln(2)+\left({1-x+x^2}\right)^{-2}\Bigg(
\left({83\over 24}-{125\over 24}x+{13\over 4}x^2+{19\over 24}x^3
-{25\over 24}x^4\right)
\nn \\
&&
\times\pi^2
+\left(-9+{43\over 2}x-34x^2+22x^3-9x^4\right)\zeta(3)
+\left(-{11\over 90}- {5\over 24}x + 
{29\over 180}x^2+{23\over 180}x^3 -
{49\over 480}x^4\right)\pi^4
\nn \\
&&
+\left[-{93\over 8}+{231\over 16}x-{279\over 16}x^2+{93\over 16}x^3
+\left(-{3\over 2}+{13\over 4}x-{7\over 12}x^2-{11\over 8}x^3
\right)\pi^2+\left(12-12x+8x^2\right.\right.
\nn \\
&&
\left.-x^3\right)\zeta(3)\bigg]\ln(x)
+\left[{9\over 2}-{43\over 8}x+{17\over 8}x^2
+{29\over 8}x^3-{9\over 2}x^4+
\left({x\over 4}+{x^2\over 2}+{5\over 24}x^3
+{19\over 48}x^4\right)
\pi^2\right]\ln^2(x)
\nn \\
&&
+\left({67\over 24}x-{5\over 4}x^2-{2\over 3}x^3\right)\ln^3(x)
+\left({7\over 48}x+{5\over 96}x^2-{x^3\over 12}
+{43\over 96}x^4\right)\ln^4(x)
+\Bigg\{3x+3x^3+\left({7\over 6}x\right.
\nn \\
&&
\left.-{73\over 24}x^2+{15\over 8}x^3\right)\pi^2
+\left(-6+6x-x^2-4x^3\right)\zeta(3)
+\left[-8+{21\over 2}x-{45\over 4}x^2+x^4 
+\left(1-{x\over 6}+{x^2\over 12}
\right.\right.
\nn \\
&&
\left.\left.
-{x^3\over 3}
-{x^4\over 8}\right)\pi^2\right]\ln(x)
+\left(6-11x+{35\over 4}x^2-{15\over 8}x^3\right)\ln^2(x)
+\left({2\over 3}+{x\over 12}-{x^3\over 3}
+{5\over 24}x^4\right)\ln^3(x)\Bigg\}
\nn \\
&&
\times\ln(1-x)+\bigg[{7\over 2}
-6x+{45\over 4}x^2-6x^3+{7\over 2}x^4
+\left(-{17\over 24}+{7\over 6}x-{25\over 24}x^2
-{13\over 48}x^4\right)\pi^2
+\left(-3+{23\over 4}x
\right.
\nn \\
&&
\left.
-{23\over 4}x^2+{9\over 8}x^3\right)\ln(x)
+\left({7\over 2}-{41\over 8}x+{31\over 8}x^2\right.
\left.+{3\over 8}x^3
-{13\over 16}x^4\right)\ln^2(x)\bigg]\ln^2(1-x)
+\left[{3\over 8}x+{1\over 6}x^2
\right.
\nn \\
&&
+{3\over 8}x^3
+\left(-4+{29\over 6}x-{49\over 12}x^2+{5\over 6}x^3
+{7\over 8}x^4\right)\ln(x)\bigg]\ln^3(1-x)
+\left({1\over 32}-{3\over 4}x+{71\over 48}x^2
-{29\over 24}x^3\right.
\nn \\
&&
\left.
+{9\over 32}x^4\right)
\ln^4\left(1-x\right)
+\Bigg\{8-16x+24x^2-16x^3
+8x^4
+\left({7\over 3}-3x+{3\over 4}x^2
+{5\over 6}x^3-{2\over 3}x^4\right)\pi^2
\nn \\
&&
+\left[-6+{11\over 2}x-4x^2+x^3+\left(2
-{11\over 4}x+{7\over 4}x^2
+{x^3\over 4}
-x^4\right)\ln(x)\right]\ln(x)
+\left[{3\over 2}x-{x^2\over 4}+x^3\right.
\nn \\
&&
+\left(-4+9x-{15\over 2}x^2+2x^3\right)\ln(x)
+\left(-1-{7\over 2}x+{25 \over 4}x^2
-5x^3+2x^4\right)\ln(1-x)\Bigg]\ln(1-x)
+\Big(2
\nn \\
&&
\left.-4x+6x^2-4x^3+2x^4\right)
{\rm Li}_2(x)\Bigg\}{\rm Li}_2\left(x\right)
+\Bigg\{-8+16x-24x^2+16x^3-8x^4 
+\left[-{2\over 3}+{4\over 3}x\right.
\nn \\
&&
\left.+{x^2\over 2}-{5\over 3}x^3
+{2\over 3}x^4\right]\pi^2
+\left[6-8x+9x^2-3x^3+\left({3\over 2}x-{x^2\over 2}
-{3\over 2}x^3+2x^4\right)\ln(x)\right]\ln(x)+\Bigg[-x
\nn \\
&&
\left.-{x^2\over 4}-{x^3\over 2}+\left(10-14x+9x^2\right)\ln(x)
+\left(-8+11x-{31\over 4}x^2
+{x^3\over 2}+x^4\right)\ln(1-x)\right]\ln(1-x)
\nn \\
&&
+\left(-4+8x-12x^2+8x^3
-4x^4\right){\rm Li}_2(x)
+\left(2-4x+6x^2-4x^3+2x^4\right){\rm Li}_2(1-x)\Bigg\}
{\rm Li}_2\left(1-x\right)
\nn \\
&&
+
\left[{5\over 2}x-5x^2+2x^3+\left(-4-x+x^2+2x^3
-2x^4\right)\ln(x)+(6-6x
+x^2+4x^3)\ln(1-x)\right]{\rm Li}_3\left(x\right)
\nn \\
&&
+\left[{x\over 2}-{x^3\over 2}+(-6+5x+3x^2-5x^3)\ln(x)
+\left(6-10x+10x^3-6x^4\right)\ln(1-x)\right]
{\rm Li}_3\left(1-x\right)
\nn\\
&&
+\left(-2+{17\over 2}x-{17\over 2}x^3
+2x^4\right){\rm Li}_4\left(x\right)
+\left(7x-{9\over 2}x^2-4x^3+6x^4\right)
{\rm Li}_4\left(1-x\right)+\bigg(-6+4x
\nn\\
&&
\left.+{9\over 2}x^2-7x^3\right){\rm Li}_4\left(-{x\over 1-x}\right)
\Bigg)\,,
\label{result}
\eea
}
where ${\cal  L}_{\vep}=\left[1-\ln\left(x/(1-x)\right)\right]
\ln\left(\vep_{cut}/\vep\right)$.


\begin{thebibliography}{99}

\bibitem{Jad} S. Jadach {\it et
              al.} in G. Altarelli, T. Sj\"ostrand and F. Zwirner (eds.),
              {\it Physics at LEP2}, CERN-96-01, hep-ph/9602393.

\bibitem{MNP} G. Montagna, O. Nicrosini, and F. Piccinini, 
              Riv.\ Nuovo\ Cim.\ {\bf 21N9}, 1 (1998).

\bibitem{Too} N. Toomi, J. Fujimoto, S. Kawabata, Y. Kurihara, and 
              T. Watanabe, Phys.\ Lett.\ B {\bf 429}, 162 (1998).

\bibitem{Heu} R.D. Heuer, D. Miller, F.Richard, and P.M. Zerwas,
              (eds.), {\it TESLA Technical design report.
              Pt. 3: Physics at an $e^+e^-$ linear collider}, 
              Report No. DESY-01-011C and  hep-ph/0106315.
 
\bibitem{Car} C. M. Carloni Calame, C. Lunardini,  G. Montagna,
              O. Nicrosini, and F. Piccinini,  
              Nucl.\ Phys.\  {\bf B584}, 459 (2000). 

\bibitem{Ber}  F.A. Berends and R. Kleiss,  Nucl.\ Phys.\  B {\bf 228},
               573 (1983).

\bibitem{CGR}  M. Caffo, R. Gatto, and E. Remiddi, Nucl.\ Phys.\  B {\bf 252},
               378 (1985). 

\bibitem{BDG}  Z. Bern, L. Dixon, and A. Ghinculov, Phys.\ Rev.\  
               D {\bf 63},  053007 (2001).

\bibitem{Jad2} S. Jadach,  W. Placzek, E. Richter-W\c{a}s, B.F.L. Ward,
               Z. W\c{a}s, Comput.\ Phys.\ Commun.\ {\bf 102}, 229 (1997).

\bibitem{JPW}  S. Jadach, W. Placzek, and B.F.L. Ward, 
               Phys.\ Lett.\ B {\bf 390}, 298 (1997).

\bibitem{Arb4} A.B. Arbuzov, G.V. Fedotovich, F.V. Ignatov, E.A. Kuraev,
               A.L. Sibidanov, Report No. BUDKER-INP-2004-70,  
               and hep-ph/0504233.

\bibitem{BFMR} R. Bonciani, A. Ferroglia, P. Mastrolia, and  E. Remiddi,
               Nucl.\ Phys.\  {\bf B701}, 121 (2004). 

\bibitem{Bon}  R. Bonciani, A. Ferroglia, P. Mastrolia, E. Remiddi,
               and J.J. van der Bij,  Nucl.\ Phys.\ {\bf B716},  280 (2005).

\bibitem{Smi}  V.A. Smirnov,  Phys.\ Lett.\ B {\bf 524}, 129 (2002);
               Nucl.\ Phys.\ Proc.\ Suppl.\  {\bf 135}, 252 (2004).

\bibitem{HeiSmi} G. Heinrich and  V.A. Smirnov, 
                 Phys.\ Lett.\ B {\bf 598}, 55 (2004).

\bibitem{Cza}  M. Czakon, J. Gluza, and T. Riemann,
               Nucl.\ Phys.\ Proc.\ Suppl.\ {\bf 135}, 83 (2004);
               Phys.\ Rev.\  D {\bf 71}, 073009 (2005).

\bibitem{BonFer}  R. Bonciani and  A. Ferroglia,
                  Phys.\ Rev.\ D {\bf 72}, 056004 (2005).

\bibitem{AKS}  A.B. Arbuzov, E.A. Kuraev, and B.G. Shaikhatdenov, 
               Mod.\  Phys.\ Lett.\ A {13}, 2305 (1998).

\bibitem{GTB}  E.W. Glover, J.B. Tausk, and J.J. van der Bij, 
               Phys.\ Lett.\  B {\bf 516}, 33 (2001).

\bibitem{Fad2} V.S. Fadin, E.A. Kuraev, L.N. Lipatov,
               N.P. Merenkov, and L. Trentadue,  
               Fhys.\ Atom.\ Nucl.\  {\bf 56}, 1537 (1993)
               [Yad.\ Fiz.\ {\bf 56} 145 (1993)].

\bibitem{Arb}  A.B. Arbuzov, V.S. Fadin, E.A. Kuraev, L.N. Lipatov,
               N.P. Merenkov, and L. Trentadue,  
               Nucl.\ Phys.\  {\bf B485}, 457 (1997).

\bibitem{Jad1} S. Jadach, M. Melles, B.F.L. Ward, and S.A. Yost, Phys.\
               Lett.\ B {\bf 450}, 262 (1999).

\bibitem{Pen}  A.A. Penin,  Phys.\ Rev.\ Lett.\ {\bf 95}, 010408 (2005).

\bibitem{Kin}  T. Kinoshita, J.\ Math.\ Phys.\  {\bf 3}, 650, (1962). 

\bibitem{LeeNau} T.D. Lee and M. Nauenberg,  
                 Phys.\ Rev.\ B {\bf 133},  1549 (1964).

\bibitem{BenSmi} M. Beneke and  V.A. Smirnov, Nucl.\ Phys.\  {\bf B522}, 321 (1998).

\bibitem{Smi1} V.A. Smirnov, {\it Applied Asymptotic Expansions in Momenta and Masses}
               (Springer-Verlag, Heidelberg, 2001).

\bibitem{SteWei} G. Sterman and  S. Weinberg, 
                 Phys.\ Rev.\ Lett.\ {\bf 39}, 1436 (1977).

\bibitem{FKPS} B. Feucht, J.H. K\"uhn, A.A. Penin, and V.A. Smirnov,
               Phys.\ Rev.\ Lett.\ {\bf 93}, 101802 (2004).

\bibitem{Sud}  V.V. Sudakov, Zh.\ Eksp.\ Teor.\ Fiz.\ {\bf 30}, 87 (1956).

\bibitem{YFS}  D.R. Yennie, S.C. Frautschi, and H. Suura, Ann. Phys. 
               {\bf 13}, 379 (1961).

\bibitem{Jac}  R. Jackiw, Ann.\ Phys.\ {\bf 48}, 292 (1968); {\bf 51},
               575(1969).

\bibitem{Mue}  A.H. Mueller, Phys.\ Rev.\ D {\bf 20}, 2037 (1979).

\bibitem{Col}  J.C. Collins, Phys.\ Rev.\ D {\bf 22}, 1478 (1980).

\bibitem{Sen}  A. Sen, Phys.\ Rev.\  D {\bf 24},  3281 (1981); {\bf 28},
               860 (1983).

\bibitem{FreTay} J. Frenkel and J.C. Taylor, 
                 Nucl.\ Phys.\ {\bf B116},  185 (1976).        

\bibitem{JKPS} B. Jantzen, J.H. K\"uhn, A.A. Penin, and V.A. Smirnov,
               Phys.\ Rev.\ D {\bf 72}, 051301(R) (2005).

\bibitem{KraLam} G. Kramer and B. Lampe,  
                 {Z. Phys.} C {\bf 34}, 497 (1987); {\bf 42}, 504(E)
                 (1989).

\bibitem{MMN}  T. Matsuura, S.C. van der Marck, and W.L. van Neerven,
               Nucl.\ Phys.\ {\bf B319}, 570 (1989).

\bibitem{Bur}  G.J.H. Burgers, Phys.\ Lett.\ B {\bf 164}, 167 (1985).

\bibitem{BMR}  R. Barbieri, J.A. Mignaco, and E. Remiddi, 
               Nuovo\ Cim.\  A {\bf 11}, 824 (1972).

\bibitem{MasRem} P. Mastrolia and E. Remiddi, Nucl.\ Phys.\ {\bf B664}, 341 (2003). 

\bibitem{MVV}  S. Moch, J.A.M. Vermaseren, and  A. Vogt. Report No.
               DESY-05-106, and hep-ph/0507039.

\bibitem{Ste}  G. Sterman, Nucl.\ Phys.\  {\bf B281}, 310 (1987).


\bibitem{KPSSS} B.A. Kniehl, A.A. Penin, Y. Schroder, V.A. Smirnov, 
                and M. Steinhauser,
                Phys.\ Lett.\ B {\bf 607}, 96 (2005).

\bibitem{Cat}  S. Catani, Phys.\ Lett.\ B {\bf 427}, 161 (1998).

\bibitem{Boh}  M. B\"ohm, A. Denner, and W. Hollik, 
               Nucl.\ Phys.\  {\bf B304}, 687 (1988).

\bibitem{Jeg}  F. Jegerlehner, J.\ Phys.\ G {\bf 29}, 101 (2003).

\bibitem{Arb1} A.B. Arbuzov, E.A. Kuraev, 
               N.P. Merenkov, and L. Trentadue, Nucl.\ Phys.\ 
               {\bf B474}, 271 (1996); 
               Phys.\ Atom.\ Nucl.\  {\bf 60}, 591 (1997)
               [Yad.\ Fiz.\ {\bf 60} 673 (1997)]. 

\bibitem{Fad}  V.S. Fadin, L.N. Lipatov, A.D. Martin and M. Melles,
               Phys.\ Rev.\ D {\bf 61}, 094002 (2000).

\bibitem{KPS}  J.H. K\"uhn, A.A. Penin, and V.A. Smirnov,
               Eur.\ Phys.\ J.\   C {\bf 17}, 97 (2000);
               Nucl.\ Phys.\ Proc.\ Suppl.\ {\bf 89}, 94 (2000).

\bibitem{KMPS} J.H. K\"uhn, S. Moch, A.A. Penin, and V.A. Smirnov,
               Nucl.\ Phys.\  {\bf B616}, 286 (2001);
               {\bf B648}, 455(E) (2002).
          


%
%




\end{thebibliography}
\end{document}